\documentclass[aps,preprint,floatfix,nofootinbib,showpacs]{revtex4}
\usepackage{graphicx}

\newcommand{\U}{{\cal U}}
\begin{document}



\title{Higgs Boson Decays into Single Photon plus Unparticle}

\renewcommand{\thefootnote}{\fnsymbol{footnote}}

\author{ 
Kingman Cheung$^{1,2}$, 
Chong Sheng Li$^3$
and Tzu-Chiang Yuan$^{2}$
 }
\affiliation{$^1$Department of Physics, National Tsing Hua University, 
Hsinchu 300, Taiwan
\\
$^2$Physics Division, National Center for Theoretical Sciences,
Hsinchu 300, Taiwan
\\
$^3$Department of Physics, Peking University, Beijing 100871, China
}

\renewcommand{\thefootnote}{\arabic{footnote}}
\date{\today}

\begin{abstract}
The decay of the standard model Higgs boson into a single photon and a
vector unparticle through a one-loop process is studied.  For an
intermediate mass Higgs boson, this single photon plus unparticle mode can
have a branching ratio comparable with the two-photon discovery mode.
The emitted photon has a continuous energy spectrum encoding the
nature of the recoil unparticle.  It can be measured in precision
studies of the Higgs boson after its discovery.
\end{abstract}

\pacs{14.80.-j, 12.38.Qk, 12.90.+b, 13.40.Em}

\maketitle

\section{Introduction}

The electroweak-symmetry breaking of the standard model (SM) in
particle physics that provides masses to its particle contents will
soon be tested at the CERN Large Hadron Collider (LHC), which is
scheduled to be online in later part of the year 2008.  The simplest
version of the electroweak-symmetry breaking consists of an elementary
scalar boson known as the Higgs boson $H$.  On the theoretical side, the
best fit value of electroweak precision data for the Higgs boson mass is
$m_H = 89^{+38}_{-28}$ 
GeV with a 95\% CL upper limit of $m_H < 144$
GeV \cite{pdg}, while the direct searches at LEP puts a lower limit on
$m_H > 114.5$ GeV \cite{lep}.  For an intermediate-mass Higgs boson in
the mass range of $115-140$ GeV, the best hope to search for it at the
LHC is the two-photon mode \cite{higgshunters} even though the
branching ratio for the two-photon decay mode is only $10^{-3}$.  The
Higgs boson will manifest as a sharp peak standing above the continuum
background in the diphoton invariant mass spectrum.  The position of
the peak indicates the mass of the Higgs boson.
In the rest frame of the Higgs boson, the energy of the photon will be
exactly one half of the Higgs boson mass.  
Another rare decay mode of the
Higgs boson into a single photon $\gamma$ and the massive 
neutral gauge boson $Z$ of the SM
has a branching ratio of order $10^{-3}$, in which
the photon is also mono-energetic in the rest frame of the Higgs boson.

In this work, we point out a possible rare decay mode of the Higgs
boson in the scheme of unparticle proposed in \cite{unparticle}.  The Higgs
boson can decay into a single photon plus a vector unparticle $\U$. The
salient feature of this decay mode is that the photon energy has a
continuous spectrum in the rest frame of the Higgs boson, in contrast
to $H \to \gamma\gamma$ and $\gamma Z$.  Therefore, by measuring the
photon energy spectrum in the Higgs boson decay, one can discriminate
the presence of the unparticle or not.
Note that we cannot use $H\to \gamma {\cal U}$ for the discovery of
the Higgs boson, because of the missing energy carried away by the
unparticle.  Therefore, the decay mode that we propose in this work
will be in the precision studies of the Higgs boson decay.  Perhaps, it
can be done at the future International Linear Collider.

The notion of unparticle was introduced in \cite{unparticle} to
describe a possible scale-invariant hidden sector that possesses an
infrared fixed point at a higher scale $\Lambda_\U$ presumably above
the Fermi scale.
Such a hidden sector is assumed interacting with the visible SM sector weakly enough 
to be describable by an effective field theory governed by non-renormalizable operators 
suppressed by inverse powers of $\Lambda_\U$.
Phenomenological implications of unparticle have since been studied 
by many authors \cite{
unparticle-propagator,
CKY,
Luo-Zhu,
Chen-Geng,  
Ding-Yan,
Liao-1,
Aliev-Cornell-Gaur,
Catterall-Sannio,
Li-Wei,
Lu-Wang-Wang,
Fox-Rajaraman-Shirman,
Greiner,
Davoudiasl,  
Choudhury-Ghosh-Mamta,
Chen-He,        
Mathews-Ravindran,
Zhou,
Liao-Liu,
minimal-walking,
Bander-Feng-Rajaraman-Shirman,
Rizzo,
ungravity,
Chen-He-Tsai,
Zwicky,
Kikuchi-Okada,
Mohanta-Giri,
Huang-Wu,
Lenz,
Choudhury-Ghosh,
Zhang-Li-Li,
Nakayama,
Desh-He-Jiang,
BGHNS, 
Delgado-Espinosa-Quiros,
colored-unparticle,
Neubert,
Hannestad-Raffelt-Wong,
HMS,
Desh-Hsu-Jiang,
Kumar-Das,
BCG,
Liao-2,
Majumdar,
Alan-Pak-Senol,
Freitas-Wyler,
GOS,
Hur-Ko-Wu,
Anchordoqui-Goldberg,
Majhi, 
McDonald,
KMRT,
DMR,
Kobakhidze,
Balantekin-Ozansoy,
Aliev-Savci,
LKQWL,  
Iltan, 
CHLTW,
Lewis, 
AKTVV,
CHHL, 
Sahin-Sahin},
while more conceptual aspects of unparticle were
explored by others 
\cite{Stephanov,Krasnikov,HEIDI,Ryttov-Sannino,JPLee}.
Due to the exact scale invariance, unparticle does not behave like
ordinary particles.  It has a continuous spectral density and behaves
like a collection of $d_\U$ massless particles, where $d_\U$ is the
scaling dimension of the unparticle operator ${\cal O}_\U$.
Thus unparticle does not have a definite mass and 
is just like the massless photon that it has no rest frame.
This implies that a real unparticle is stable and cannot decay.
Direct signals of unparticle can nevertheless be detected
in the missing energy and momentum distribution carried away by the
unparticle once it was produced in a process \cite{unparticle}, while 
virtual unparticle effect can be probed via its interference  
with the SM amplitudes \cite{unparticle-propagator,CKY}.
Thus, even in the case of 2-body decay like
$H \to \gamma \U$,
the energy spectrum of the photon is no longer a delta function peaked at $m_H/2$ but
rather a continuous one spreading from zero to $m_H/2$.

One more remark before we come to the details of the calculation is
that $H \to \gamma {\cal U}$ is a one-loop process, in analogous to $H
\to \gamma Z$, but with only SM fermions flowing in the loop.  The
major contribution comes from the top-quark loop.  The vector coupling
of the top quark to the unparticle can be parameterized by
$(\lambda^t_1/\Lambda_\U^{d_\U -1}) \bar t \gamma_\mu t\,
O_{\U}^{\mu}$.  As long as the coupling is flavor dependent, the
constraint coming from the top quark physics
\footnote{We note that strong constraints for flavor independent
couplings between SM fermions and unparticle operators have been
obtained in \cite{Choudhury-Ghosh} using the $t \bar t$ production
cross section from the Tevatron.} is rather loose, even though the
constraints for other fermions (e.g. electron) are very stringent
\cite{CKY}.  Therefore, we are allowed to use $\lambda^t_1 \sim 1$ and
$\Lambda_\U \sim 1$ TeV for the top quark without upsetting existing
constraints.

\section{Unparticle calculation}

The interaction of a vector unparticle $\U$ with a standard model fermion $f$ 
is given by \cite{unparticle,CKY}
\begin{eqnarray}
{\cal L}_{\mathrm{eff}} & \ni & \frac{1}{\Lambda_\U^{d_\U-1}} \bar f \left( 
\lambda_1^f \gamma_\mu + \lambda^{\prime f}_1 \gamma_\mu \gamma_5 
\right) f O^\mu_{\U} 
\end{eqnarray}
where $ \lambda_1^f $ and $ \lambda_1^{\prime f}$ are the unknown
vector and axial vector couplings.  The process $H \to \gamma \U$ can
be induced at one-loop level with the standard model fermions circling the loop.
The most dominant contribution comes from the top-quark loop.
We note that the $W$ boson loop that contributes
significantly in the two-photon mode does not contribute in the
unparticle case.
The Lorentz-invariant decay amplitude $\cal M$ for $H \to \gamma(k) +
\U(P_\U)$ is dictated by gauge invariance of electromagnetism and can
be written as
\begin{eqnarray}
{\cal M} & = & \epsilon^*_\mu(k, \lambda) \epsilon^*_\nu(P_\U, \lambda^\prime)
 {\cal M}^{\mu \nu}
\end{eqnarray}
with
\begin{eqnarray}
{\cal M}^{\mu\nu} & = & \left( P^\mu_\U k^\nu - g^{\mu \nu} P_\U \cdot k 
\right) {\cal A} \; .
\end{eqnarray}
The loop-induced amplitude $\cal A$ can be extracted from previous 
$H \gamma Z$ calculations \cite{Cahn-Chanowitz-Fleishon,Bergstrom-Hulth,Gamberini-Giudice-Ridolfi,Gunion-Kane-Wudka,Weiler-Yuan} by 
the following substitutions 
\begin{eqnarray}
- \frac{g}{\cos \theta_w} \left( \frac{1}{2} T^{3L}_f - Q_f \sin^2 
\theta_w \right)
& \longrightarrow & \frac{\lambda_1^f}{\Lambda^{d_\U - 1}_\U} \;\; , \\
m_Z^2 & \longrightarrow & P^2_\U \;\; .
\end{eqnarray}
As in the $H\gamma Z$ case, the axial vector coupling $\lambda_1^{\prime f}$
does not contribute to $\cal A$ as it is forbidden by charge conjugation.
Thus
\begin{eqnarray}
{\cal A} & = & \frac{\alpha}{\pi m_W \Lambda^{d_\U - 1}_\U} A_F
\end{eqnarray}
where
\begin{eqnarray}
A_F & = & \sum_f N_c^f \frac{\lambda^f_1 Q_f}{\sin \theta_w} 
\left[ 
I \left( x_f, y_f \right) - J \left( x_f, y_f \right) 
\right] 
\end{eqnarray}
with $x_f = 4 m_f^2/m_H^2$, $y_f = 4 m_f^2 / P^2_\U$ and $m_f$ is a fermion mass. 
The loop functions $I(x,y)$ and $J(x,y)$ are given by 
\begin{eqnarray}
I \left(x, y \right) & = &
\frac{xy}{\left(  x - y \right)} \left[ \frac{1}{2} -  J\left( x, y \right)
+  \frac{x}{\left( x - y \right)} \left[ g\left( x \right) - g \left( y \right) \right] \right] \;\; , \\
J \left( x, y \right) & = & - \frac{x y }{2 \left( x - y \right)} \left[ f \left( x \right) - f \left( y \right) \right]
\end{eqnarray}
with
\begin{eqnarray}
f\left( x \right) & = &
\left\{ 
\begin{array}{ll}
\left[
\sin^{-1} \left( \frac{1}{\sqrt{x}} \right) \right]^2 \quad & \quad \mbox{if} \quad x \ge 1 \\
-\frac{1}{4} \left[ \ln \left( \frac{1 + \sqrt{1-x}}{1 - \sqrt{1-x}}\right) - i \pi \right]^2 \quad & \quad \mbox{if} \quad x < 1
\end{array}
\right. \;\; , \\
g\left( x \right) & = &
\left\{ 
\begin{array}{ll}
\sqrt{x - 1}
\sin^{-1} \left( \frac{1}{\sqrt{x}} \right)  \quad & \quad \mbox{if} \quad x \ge 1 \\
\frac{1}{2} \sqrt{1-x} \left[ \ln \left( \frac{1 + \sqrt{1-x}}{1 - \sqrt{1-x}}\right) - i \pi \right] \quad & \quad \mbox{if} \quad x < 1 \end{array}
\right. \;\; .
\end{eqnarray}
The energy distribution of the emitted photon for this process can be
easily derived as
\begin{eqnarray}
\frac{d \Gamma}{d E_\gamma} & = &
\frac{\alpha^2}{4 \pi^4 m_W^2} A_{d_\U} m_H E_\gamma^3 \frac{1}{\Lambda^2_\U} 
\left( \frac{P^2_\U}{\Lambda^2_\U} \right)^{d_\U - 2}  \vert A_F \vert^2
\end{eqnarray}
with  $P^2_\U = m_H^2 - 2 m_H E_\gamma$ and $E_\gamma$ lies in the range $\left[ 0, m_H/2 \right]$. 
$A_{d_\U}$ is the normalization for the unparticle phase space \cite{unparticle}
\begin{eqnarray}
A_{d_\U} & = & \frac{16 \pi^{\frac{5}{2}}} {\left( 2 \pi \right)^{2 d_\U} } 
\frac{\Gamma \left( d_\U + \frac{1}{2} \right)} { \Gamma\left( d_\U - 1 \right) \Gamma \left( 2 d_\U \right) } \; \; .
\end{eqnarray}

\section{Results}

\begin{figure}[t!]
\centering
\includegraphics[width=4.5in]{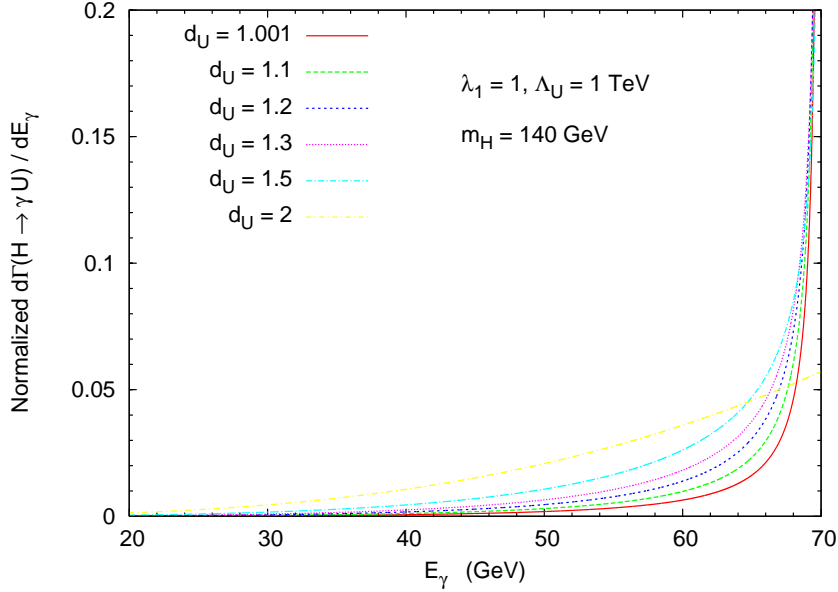}
\caption{
\label{NormalizedDist} \small
Normalized photon energy spectrum for different values 
of $d_\U=1.001,\,1.1,\,1.2,\,1.3,\,1.5$ and 2.
}
\end{figure}

In Fig.~\ref{NormalizedDist}, we plot the normalized energy spectrum
of the emitted photon from $H \to \gamma \U$ for various values of the
scaling dimension $d_\U=1.001,\,1.1,\,1.2,\,1.3,\,1.5$ and 2 with a
Higgs boson mass of 140 GeV.  As the scaling dimension approaches unity,
the distribution becomes a delta function peaked at one half of the
Higgs boson mass.  As $d_\U$ moves away from unity, the energy spectrum
begins to flatten out gradually.  For simplicity, we have only included
the top quark in the loop, because it is the most dominant and the size 
of the coupling that we used is consistent with the top quark physics.
Even if we include all SM fermions in the loop, there is hardly 
any visible change to the figure.

\begin{figure}[t!]
\centering
\includegraphics[width=4.5in]{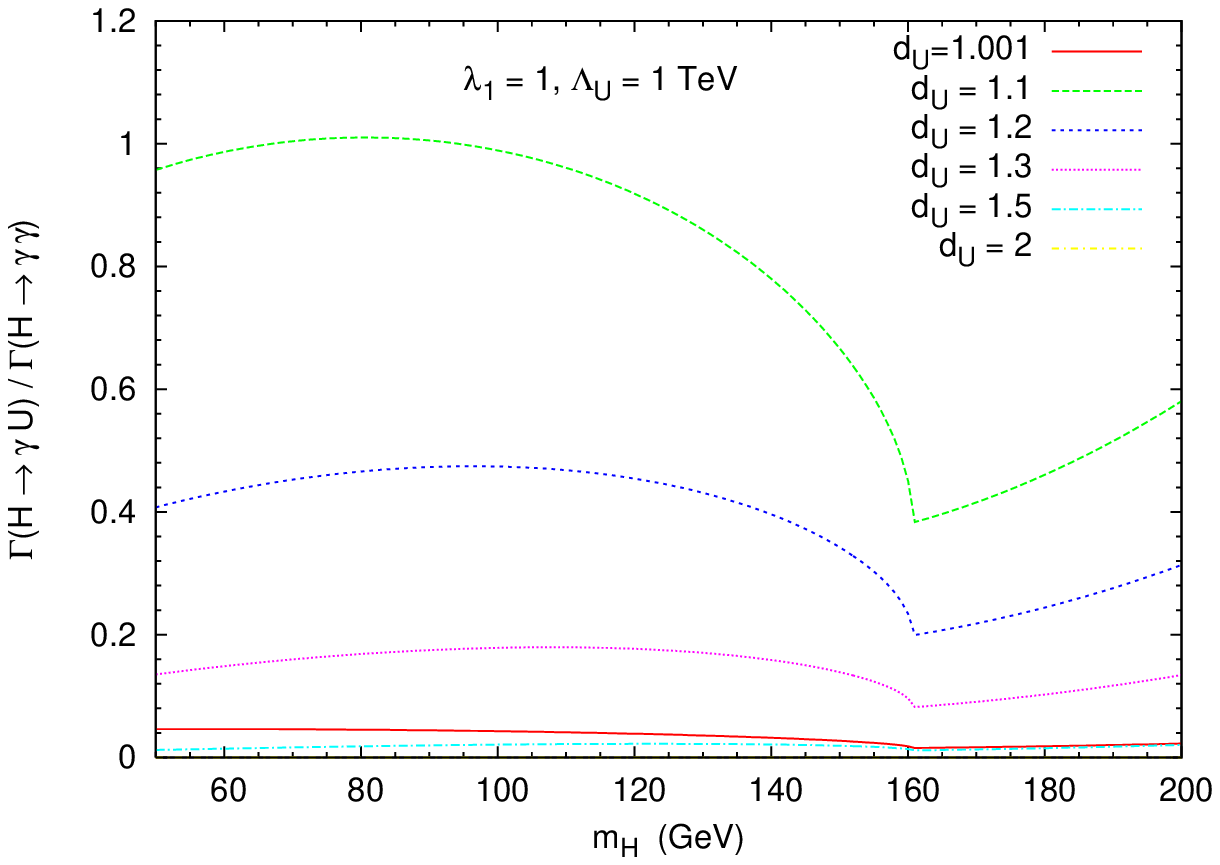}
\caption{
\label{Branching1} \small
Branching ratio $\Gamma \left( H \to \gamma \U\right) / \Gamma \left(
H \to \gamma \gamma \right)$ for different values of $d_{\cal
U}=1.001,\,1.1,\,1.2,\,1.3,\,1.5$ and 2.  }
\end{figure}
In Fig.~\ref{Branching1}, we compare the decay rate of the single
photon mode $H \to \gamma \U$ with that of the two-photon mode $H \to
\gamma \gamma$. One can see that for $d_\U = 1.1$ and 50 GeV $< m_H <$
100 GeV, both modes can have the same branching ratio.

\begin{figure}[t!]
\centering
\includegraphics[width=5in]{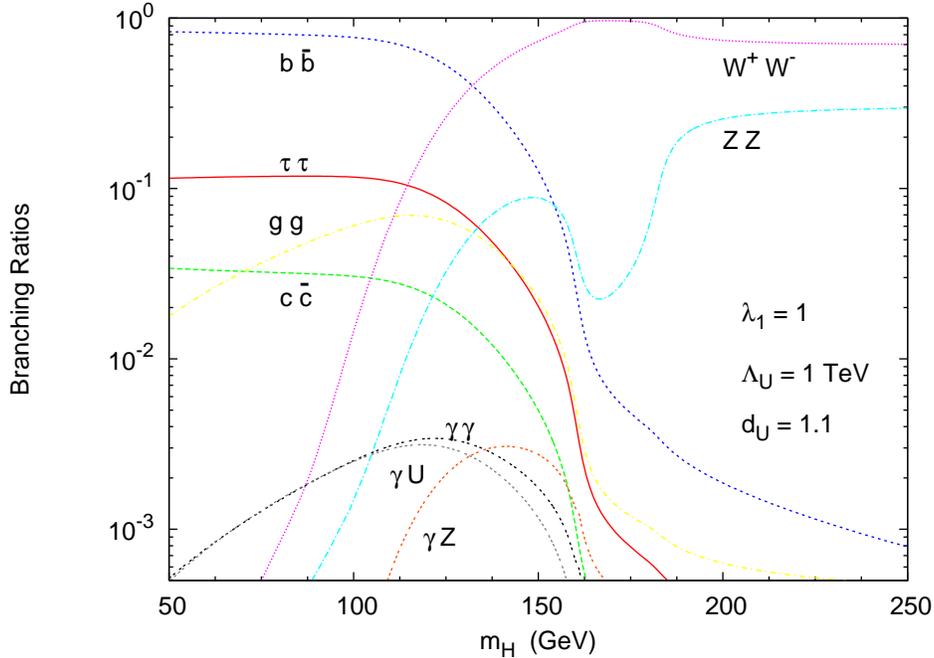}
\caption{
\label{Branching2} \small
Branching ratios of various decay modes of the Higgs boson versus
the mass of the Higgs boson for $\lambda^t_1 =1$, $\Lambda_\U = 1$ TeV
and $d_\U = 1.1$.}
\end{figure}
In Fig.~\ref{Branching2}, 
the branching ratios of various Higgs boson decay modes
are plotted as a function of the Higgs boson mass including the process that
we study in this paper.  We have used the running masses for all the 
fermions to account for the QCD radiative corrections as well as the off-shell
decay formulae in the $WW$, $ZZ$ and $t\bar t$ modes.  It is clear from the
figure that the $\gamma\U$ mode is comparable to the $\gamma\gamma$
mode and larger than the $\gamma Z$ mode for all the Higgs boson mass range 
up to 130 GeV.  

In summary, we have studied the Higgs boson decay mode into a single
photon plus a vector unparticle.  This mode can be used to probe the
hidden unparticle sector since the emitted energy of the single photon is
encoded with the information of missing energy of the recoil
unparticle. This mode is particularly useful for an intermediate mass
Higgs boson since it may have comparable branching ratio with the two-photon
discovery mode of the Higgs boson decay. 
Finally, it should be of interest to extend the present 
study to the case of the
tensor unparticle operator as well as the $HZ\U$ vertex.

\section*{Acknowledgments}
This research was supported in parts by the NSC
under Grant No.\ NSC 96-2628-M-007-002-MY3, the NCTS of Taiwan (Hsinchu),
and the
National Natural Science Foundation of China under grants 
No.10421503, No.10575001 and No.10635030. 
One of us (TCY) would like to thank the hospitality of
the Department of Physics of Peking University where this work was initiated.
The authors would also like to thank W. Y. Keung for useful communications.


\begin{thebibliography}{10}

\bibitem{pdg}
Particle Data Group, Review of Particle Physics, J. Phys. G:
Nucl. Part. Phys. 33, 1 (2006).

\bibitem{lep}
  R.~Barate {\it et al.}  [LEP Working Group for Higgs boson searches],
  Phys.\ Lett.\ B {\bf 565}, 61 (2003), [hep-ex/0306033].
 
\bibitem{higgshunters}
J. F. Gunion, H. E. Haber, G. Kane and S. Dawson, {\it The Higgs Hunter's Guide}, Addison-Wesley, Reading (1990).
 
\bibitem{unparticle}
H.~Georgi,
Phys. Rev. Lett. {\bf 98}, 221601 (2007)
[hep-ph/0703260].

\bibitem{unparticle-propagator}
H.~Georgi,
Phys. Lett. B {\bf 650}, 275 (2007), arXiv:0704.2457 [hep-ph]. 

\bibitem{CKY}
K.~Cheung, W.~Y.~Keung and T.~C.~Yuan,
Phys. Rev. Lett. {\bf 99}, 051803 (2007), 
arXiv:0704.2588 [hep-ph];
Phys. Rev. D {\bf 73}, 075015 (2007), arXiv:0706.3155 [hep-ph]. 

\bibitem{Luo-Zhu}
M.-x. Luo and G.-h. Zhu,
arXiv:0704.3532 [hep-ph];
arXiv:0708.0671 [hep-ph].
    
\bibitem{Chen-Geng}  
C.-H. Chen and C.-Q. Geng,
Phys. Rev. D {\bf 76}, 036007 (2007), arXiv:0706.0850 [hep-ph]; 
arXiv:0709.0235 [hep-ph]; arXiv:0705.0689 [hep-ph].
    
\bibitem{Ding-Yan}
G.-J. Ding and M.-L. Yan,
Phys. Rev. D {\bf 76}, 075005 (2007), arXiv:0705.0794 [hep-ph];
arXiv:0706.0325 [hep-ph];
arXiv:0709.3435 [hep-ph].

\bibitem{Liao-1}
Y.~Liao,
Phys. Rev. D {\bf 76}, 056006 (2007), arXiv:0705.0837 [hep-ph].
 
\bibitem{Aliev-Cornell-Gaur}
T.~M. Aliev, A.~S. Cornell and N.~Gaur,
JHEP {\bf 0707}, 072 (2007),
arXiv:0705.4542 [hep-ph];
arXiv:0705.1326 [hep-ph].

\bibitem{Catterall-Sannio}
S. Catterall and F. Sannino,
Phys. Rev. D {\bf 76}, 034504 (2007), arXiv:0705.1664 [hep-lat].
   
\bibitem{Li-Wei}
X.-Q. Li and Z.-T. Wei,  
Phys. Lett. B {\bf 651}, 380 (2007), arXiv:0705.1821 [hep-ph].
  
\bibitem{Lu-Wang-Wang}
C.-D. Lu, W. Wang and Y.-M. Wang,
Phys. Rev. D {\bf 76}, 077701 (2007), arXiv:0705.2909 [hep-ph].
 
\bibitem{Fox-Rajaraman-Shirman}
P. J. Fox, A. Rajaraman and Y. Shirman,
Phys. Rev. D {\bf 76}, 075004 (2007), arXiv:0705.3092 [hep-ph].

\bibitem{Greiner}
N.~Greiner,
Phys. Lett. B {\bf 653}, 75 (2007), arXiv:0705.3518 [hep-ph].

\bibitem{Davoudiasl}  
H. Davoudiasl,
Phys. Rev. Lett.  {\bf 99}, 141301 (2007), arXiv:0705.3636 [hep-ph].

\bibitem{Choudhury-Ghosh-Mamta}
D. Choudhury, D. K. Ghosh and Mamta,
arXiv:0705.3637 [hep-ph]. 

\bibitem{Chen-He}        
S.-L. Chen and X.-G. He,
arXiv:0705.3946 [hep-ph].

\bibitem{Mathews-Ravindran}
P.~Mathews and V.~Ravindran,
arXiv:0705.4599 [hep-ph].

\bibitem{Zhou}
S. Zhou,
arXiv:0706.0302 [hep-ph].
 
\bibitem{Liao-Liu}
Y.~Liao and J.-Y. Liu,
arXiv:0706.1284 [hep-ph].
  
\bibitem{minimal-walking}
R.~Foadi, M.~T.~Frandsen, T.~A.~Ryttov and F.~Sannino, 
Phys. Rev. D {\bf 76}, 055005 (2007), arXiv:0706.1696 [hep-ph].
   
\bibitem{Bander-Feng-Rajaraman-Shirman}
M.~Bander, J.~L.~Feng, A.~Rajaraman and Y.~Shirman,
arXiv:0706.2677 [hep-ph].

\bibitem{Rizzo}
T.~G.~Rizzo,
JHEP {\bf 0710}, 044 (2007), arXiv:0706.3025 [hep-ph].
 
\bibitem{ungravity}
H.~Goldberg and P.~Nath,  
arXiv:0706.3898 [hep-ph].
  
\bibitem{Chen-He-Tsai}
S.-L. Chen, X.-G. He and H.~C. Tsai, 
arXiv:0707.0187 [hep-ph ].

\bibitem{Zwicky}
R.~Zwicky, 
arXiv:0707.0677 [hep-ph];
arXiv:0710.4430 [hep-ph].

\bibitem{Kikuchi-Okada}
T.~Kikuchi and N.~Okada,
arXiv:0711.1506 [hep-ph];
arXiv:0707.0893 [hep-ph].

\bibitem{Mohanta-Giri}
R.~Mohanta and A.~K.~Giri, 
Phys. Rev. D {\bf 76}, 075015 (2007), arXiv:0707.1234 [hep-ph];
Phys. Rev. D {\bf 76}, 057701 (2007), arXiv:0707.3308 [hep-ph].

\bibitem{Huang-Wu}
C.~S.~Huang and X.-H.~Wu, 
arXiv:0707.1268 [hep-ph].
  
\bibitem{Lenz}
A.~Lenz,
Phys. Rev. D {\bf 76}, 065006 (2007), arXiv:0707.1535 [hep-ph].

\bibitem{Choudhury-Ghosh}
D.~Choudhury and D.~K.~Ghosh, 
arXiv:0707.2074 [hep-ph].

\bibitem{Zhang-Li-Li}
H. Zhang, C. S. Li and  Z. Li,
arXiv:0707.2132 [hep-ph].

\bibitem{Nakayama}
Y.~Nakayama, 
Phys. Rev. D {\bf 76}, 105009 (2007), arXiv:0707.2451 [hep-ph].
 
\bibitem{Desh-He-Jiang}
N.~G.~Deshpande, X.-G.~He and J.~Jiang,
Phys. Lett. B {\bf 656}, 91 (2007), arXiv:0707.2959 [hep-ph].

\bibitem{BGHNS} 
N. Bilic, B. Guberina, R. Horvat, H. Nikolic and H. Stefancic,
arXiv:0707.3830 [gr-qc].
  
\bibitem{Delgado-Espinosa-Quiros}
A.~Delgado, J.~R.~Espinosa and M.~Quiros,
JHEP {\bf 0710}, 094 (2007), arXiv:0707.4309 [hep-ph].

\bibitem{colored-unparticle}
G. Cacciapaglia, G. Marandella and J. Terning,
arXiv:0708.0005 [hep-ph].
 
\bibitem{Neubert}
M. Neubert, arXiv:0708.0036 [hep-ph].

\bibitem{Hannestad-Raffelt-Wong}
S.~Hannestad, G.~Raffelt and Y.~Y.~Y.~Wong,
arXiv:0708.1404 [hep-ph].

\bibitem{HMS}
K.-j. Hamada, A. Minamizaki and A. Sugamoto,
arXiv:0708.2127 [hep-ph].
 
\bibitem{Desh-Hsu-Jiang}
N.~G.~Deshpande, S.~D.~H.~Hsu and J.~Jiang,
arXiv:0708.2735 [hep-ph].
 
\bibitem{Kumar-Das}
P.~K.~Das,
arXiv:0708.2812 [hep-ph].

\bibitem{BCG}
G. Bhattacharyya, D. Choudhury and D. K. Ghosh,
Phys. Lett. B {\bf 655}, 261 (2007), arXiv:0708.2835 [hep-ph].
 
\bibitem{Liao-2}
Y.~Liao,
arXiv:0708.3327 [hep-ph].
 
\bibitem{Majumdar}
D. Majumdar, arXiv:0708.3485 [hep-ph].
  
\bibitem{Alan-Pak-Senol}
A. T. Alan and N. K. Pak, arXiv:0710.4239 [hep-ph];
A. T. Alan and N. K. Pak, arXiv:0708.3802 [hep-ph].

\bibitem{Freitas-Wyler}
A. Freitas and D. Wyler,
0708.4339 [hep-ph].
 
\bibitem{GOS}
I. Gogoladze, N. Okada and Qaisar Shafi,
arXiv:0708.4405 [hep-ph].
 
 \bibitem{Hur-Ko-Wu}
T.-i.~Hur, P.~Ko and X.-H. Wu,
arXiv:0709.0629 [hep-ph].
  
\bibitem{Anchordoqui-Goldberg}
L.~Anchordoqui and H. Goldberg,
arXiv:0709.0678 [hep-ph]. 
 
\bibitem{Majhi} 
S. Majhi, arXiv:0709.1960 [hep-ph].

\bibitem{McDonald}
J. McDonald, arXiv:0709.2350 [hep-ph].

\bibitem{KMRT}
M. C. Kumar, P. Mathews, V. Ravindran and A. Tripathi,
arXiv:0709.2478 [hep-ph].

\bibitem{DMR}
S. Das, S. Mohanty and K. Rao, arXiv:0709.2583 [hep-ph].

\bibitem{Kobakhidze}
A. Kobakhidze, arXiv:0709.3782 [hep-ph].

\bibitem{Balantekin-Ozansoy}
A. B. Balantekin and K. O. Ozansoy,
arXiv:0710.0028 [hep-ph];
O. Cakir and K. O. Ozansoy, arXiv:0710.5773 [hep-ph].

\bibitem{Aliev-Savci}
T. M. Aliev and M. Savci,
arXiv:0710.1505 [hep-ph].
 
\bibitem{LKQWL}  
X. Liu, H.-W. Ke, Q.-P. Qiao, Z.-T. Wei and X.-Q. Li, arXiv:0710.2600 [hep-ph].

\bibitem{Iltan} 
E. O. Iltan,
arXiv:0710.2677 [hep-ph]; arXiv:0711.2744 [hep-ph].
 
\bibitem{CHLTW}
S.-L. Chen, X.-G. He, X.-Q. Li, H.-C. Tsai and Z.-T. Wei,
arXiv:0710.3663 [hep-ph]. 

\bibitem{Lewis} 
I. Lewis, arXiv:0710.4147 [hep-ph].

\bibitem{AKTVV}
G. L. Alberghi, A. Y. Kamenshchik, A. Tronconi, G. P. Vacca and G. Venturi,
arXiv:0710.4275 [hep-th].

\bibitem{CHHL} 
S.-L. Chen, X.-G. He, X.-P. Hu and Y. Liao,
arXiv:0710.5129 [hep-ph].

\bibitem{Sahin-Sahin}
I. Sahin and B. Sahin, arXiv:0711.1665 [hep-ph].

\bibitem{Stephanov}
M. A. Stephanov,
Phys. Rev. D {\bf 76}, 035008 (2007), arXiv:0705.3049 [hep-ph].

\bibitem{Krasnikov}
N.~V.~Krasnikov, 
arXiv:0707.1419 [hep-ph]. 

\bibitem{HEIDI}
J.~J.~van~der~Bij and S.~Dilcher,
Phys. Lett. B {\bf 655}, 183 (2007), arXiv:0707.1817 [hep-ph].

\bibitem{Ryttov-Sannino} 
T. A. Ryttov and F. Sannino,
Phys. Rev. D {\bf 76}, 105004 (2007), arXiv:0707.3166 [hep-th].
 
\bibitem{JPLee}
J.-P. Lee,
arXiv:0710.2797 [hep-ph]. 

\bibitem{Cahn-Chanowitz-Fleishon}
R. N. Cahn, M. S. Chanowitz and N. Fleishon,
Phys. Lett. B {\bf 82}, 113 (1979).

\bibitem{Bergstrom-Hulth}
L. Bergstrom and G. Hulth,
Nucl. Phys. {\bf B259}, 137 (1985).

\bibitem{Gamberini-Giudice-Ridolfi}
G. Gamberini, G. F. Giudice and G. Ridolfi,
Nucl. Phys. {\bf B292}, 237 (1987).

\bibitem{Gunion-Kane-Wudka}
J. F. Gunion, G. Kane and J. Wudka,
Nucl. Phys. {\bf B299}, 231 (1988). 

\bibitem{Weiler-Yuan}
T. J. Weiler and T. C. Yuan,
Nucl. Phys. {\bf B318}, 337 (1989).


 

 \end{thebibliography}
\end{document}